\documentclass[12pt]{article}%
\usepackage{amsmath,latexsym}
\usepackage{graphicx}
\usepackage{amsmath}
\usepackage{amsfonts}
\usepackage{amssymb}%
\begin{document}

\title{{Two diverse models of embedding class one}}
   \author{
Peter K. F. Kuhfittig*\\
\footnote{E-mail: kuhfitti@msoe.edu}
 \small Department of Mathematics, Milwaukee School of
Engineering,\\
\small Milwaukee, Wisconsin 53202-3109, USA}

\date{}
 \maketitle

\begin{abstract}\noindent
Embedding theorems have continued to be a topic
of interest in the general theory of relativity
since these help connect the classical theory to
higher-dimensional manifolds.  This paper deals
with spacetimes of embedding class one, i.e.,
spacetimes that can be embedded in a
five-dimensional flat spacetime.  These ideas
are applied to two diverse models, a complete
solution for a charged wormhole admitting a
one-parameter group of conformal motions and
a new model to explain the flat rotation curves
in spiral galaxies without the need for dark
matter.  \\

\end{abstract}

\emph{Highlights:}

$\bullet$ The embedding of curved spacetime in
   higher-dimensional flat spacetime is discussed.

$\bullet$ The focus is on two diverse models of
   embedding class one.

$\bullet$ The first model is a complete wormhole
   solution including the junction conditions.

$\bullet$ The second is a new model to explain
   the flat galactic rotation curves without
   the need for dark matter.

\section{Introduction}
\noindent
Embedding theorems have a long history in the
general theory of relativity, as exemplified
by the induced-matter theory discussed in Ref.
\cite{WP92}.  Of particular interest to us are
spacetimes of embedding class one.  Here we recall
that an $n$-dimensional Riemannian space is said
to be of class $m$ if $n+m$ is the lowest
dimension of the flat space in which the given
space can be embedded.  It is well known that the
exterior Schwarzschild solution is a Riemannian
space of class two.  Following Refs. \cite{MG17},
we assume a spherically symmetric metric
of class two that will be reduced to class one
by a suitable transformation discussed in Sec.
\ref{S:embedding}.  Other useful references
are \cite{G1, G2, MDRK}.

These ideas will be applied to two very
different models, a complete solution for a
charged wormhole admitting a one-parameter
group of conformal motions and a new model
to explain the flat rotation curves in
spiral galaxies without the need for dark
matter.
%END OF SECTION

\section{The embedding}\label{S:embedding}
\noindent
The discussion in Ref. \cite{MG17} begins with the
static and spherically symmetric line element (in
units in which $G=c=1$)
\begin{equation}\label{E:line1}
ds^{2}=e^{\nu(r)}dt^{2}-e^{\lambda(r)}dr^{2}
-r^{2}\left(d\theta^{2}+\sin^{2}\theta \,d\phi^{2}
\right),
\end{equation}
where $\lambda$ and $\nu$ are functions of the
radial coordinate $r$.  It is shown that this
metric of class two can be reduced to class one
and thereby embedded in the five-dimensional
flat spacetime
\begin{equation}\label{E:line2}
ds^{2}=-\left(dz^1\right)^2-\left(dz^2\right)^2
-\left(dz^3\right)^2-\left(dz^4\right)^2
+\left(dz^5\right)^2.
\end{equation}
This reduction is accomplished by the following
transformation:
 $z^1=r\,\text{sin}\,\theta\,\text{cos}\,\phi$, $z^2=
 r\,\text{sin}\,\theta\,\text{sin}\,\phi$,
 $z^3=r\,\text{cos}\,\theta$, $z^4=\sqrt{K}\,e^{\frac{\nu}{2}}
 \,\text{cosh}{\frac{t}{\sqrt{K}}}$, and $z^5=\sqrt{K}
 \,e^{\frac{\nu}{2}}\,\text{sinh}{\frac{t}{\sqrt{K}}}$.
The differentials of these components are
\begin{equation}
dz^1=dr\,\text{sin}\,\theta\,\text{cos}\,\phi + r\,
\text{cos}\,\theta\,\text{cos}\,\phi\,
d\theta\,-r\,\text{sin}\,\theta\,\text{sin}\,\phi\,d\phi,
\end{equation}

\begin{equation}
dz^2=dr\,\text{sin}\,\theta\,\text{sin}\,\phi + r\,
\text{cos}\,\theta\,\text{sin}\,\phi\,
d\theta\,+r\,\text{sin}\,\theta\,\text{cos}\,\phi\,d\phi,
\end{equation}

\begin{equation}
dz^3=dr\,\text{cos}\,\theta\, - r\,\text{sin}\,\theta\,d\theta,
\end{equation}

\begin{equation}
dz^4=\sqrt{K}\,e^{\frac{\nu}{2}}\,\frac{\nu'}{2}\,
\text{cosh}{\frac{t}{\sqrt{K}}}\,dr + e^{\frac{\nu}{2}}\,
\text{sinh}{\frac{t}{\sqrt{K}}}\,dt,
\end{equation}
and
\begin{equation}
dz^5=\sqrt{K}\,e^{\frac{\nu}{2}}\,\frac{\nu'}{2}\,
\text{sinh}{\frac{t}{\sqrt{K}}}\,dr + e^{\frac{\nu}{2}}\,
\text{cosh}{\frac{t}{\sqrt{K}}}\,dt,
\end{equation}
where the prime denotes differentiation with respect
to the radial coordinate $r$.  To facilitate the substitution
into Eq. (\ref{E:line2}), we first obtain the expressions
for $-\left(dz^1\right)^2-\left(dz^2\right)^2-\left(dz^3\right)^2$
and $-\left(dz^4\right)^2+\left(dz^5\right)^2$:
\begin{equation}\label{E:part1}
-\left(dz^1\right)^2-\left(dz^2\right)^2-\left(dz^3\right)^2=
-dr^2-r^{2}\left(d\theta^{2}+\sin^{2}\theta\, d\phi^{2} \right),
\end{equation}

\begin{equation}\label{E:part2}
-\left(dz^4\right)^2+\left(dz^5\right)^2= e^{\nu}dt^{2}
-\frac{K\,e^{\nu}}{4}\,{\nu'}^2\,dr^2.
\end{equation}
Substituting Eqs. (\ref{E:part1}) and (\ref{E:part2}) in
Eq. (\ref{E:line2}), we obtain the metric
\begin{equation}\label{E:line3}
ds^{2}=e^{\nu}dt^{2}-\left(\,1+\frac{K\,e^{\nu}}{4}\,
{\nu'}^2\,\right)\,dr^{2}-r^{2}\left(d\theta^{2}
+\sin^{2}\theta\, d\phi^{2} \right).
\end{equation}
So metric (\ref{E:line3}) is equivalent to metric (\ref{E:line1}) if
\begin{equation}\label{E:lambda1}
e^{\lambda}=1+\frac{K\,e^{\nu}}{4}\,{\nu'}^2,
   \quad K>0.
\end{equation}
This condition is equivalent to the condition derived
by Karmarkar \cite{kK48} in terms of the Riemann
curvature tensor components:
\[
   R_{1414}=\frac{R_{1212}R_{3434}+R_{1224}R_{1334}}
      {R_{2323}}, \quad R_{2323}\neq 0.
\]
(See Ref. \cite{pB16} for details.)
%END OF SECTION

\section{Conformal Killing vectors}\label{S:Killing}
    \label{S:Killing}
As noted above, we assume a static spherically
symmetric spacetime admitting a one-parameter
group of conformal motions, which are motions
along which the metric tensor remains invariant
up to a scale factor.  Equivalently, there
exist conformal  Killing vectors such that
\begin{equation}\label{E:Lie}
   \mathcal{L_{\xi}}g_{\mu\nu}=g_{\eta\nu}\,\xi^{\eta}
   _{\phantom{A};\mu}+g_{\mu\eta}\,\xi^{\eta}_{\phantom{A};
   \nu}=\psi(r)\,g_{\mu\nu},
\end{equation}
where the left-hand side is the Lie derivative of the
metric tensor and $\psi(r)$ is the conformal factor
\cite{MM96, BHL07}.  The metric tensor $g_{\mu\nu}$
is conformally mapped into itself along the vector
$\xi$, which generates the conformal symmetry.   This
type of symmetry has been used to describe relativistic
stellar-type objects, as discussed in  Refs.
\cite{HPa, HPb}.  Additional new geometric and
kinematical insights are described in Refs.
\cite{MS93, Ray08, fR10, fR12b, BHL08a}.
Two earlier studies assumed \emph{non-static}
conformal symmetry \cite{BHL07, BHL08a}.  Another
significant observation is that the Kerr black hole
is conformally symmetric \cite{CMS10}.

To study the effect of conformal symmetry, it is
convenient to use line element (\ref{E:line1})
with the opposite signature \cite{RRKKI, pK10}:
\begin{equation}
   ds^2=- e^{\nu(r)} dt^2+e^{\lambda(r)} dr^2
   +r^2( d\theta^2+\text{sin}^2\theta\, d\phi^2).
\end{equation}
Using this form, the Einstein field equations
become
\begin{equation}\label{E:Einstein1}
e^{-\lambda}
\left(\frac{\lambda^\prime}{r} - \frac{1}{r^2}
\right)+\frac{1}{r^2}= 8\pi \rho,
\end{equation}

\begin{equation}\label{E:Einstein2}
e^{-\lambda}
\left(\frac{1}{r^2}+\frac{\nu^\prime}{r}\right)-\frac{1}{r^2}=
8\pi p_r,
\end{equation}

\noindent and

\begin{equation}\label{E:Einstein3}
\frac{1}{2} e^{-\lambda} \left[\frac{1}{2}(\nu^\prime)^2+
\nu^{\prime\prime} -\frac{1}{2}\lambda^\prime\nu^\prime +
\frac{1}{r}({\nu^\prime- \lambda^\prime})\right] =8\pi p_t.
\end{equation}

Following Herrera and Ponce de Le\'{o}n \cite{HPa},
we can simplify the analysis by requiring that
$\xi^{\alpha}U_{\alpha}=0$, where $U_{\alpha}$ is the
four-velocity of the perfect fluid distribution, so
that fluid flow lines are mapped conformally onto
fluid flow lines.  From the assumption of spherical
symmetry, it now follows that
$\xi^0=\xi^2=\xi^3=0$ \cite{HPa}.  Eq. (\ref
{E:Lie}) then yields the following results:
\begin{equation}\label{E:sol1}
    \xi^1 \nu^\prime =\psi,
\end{equation}
\begin{equation}\label{E:sol2}
   \xi^1  = \frac{\psi r}{2},
\end{equation}
and
\begin{equation}\label{E:sol3}
  \xi^1 \lambda ^\prime+2\,\xi^1 _{\phantom{1},1}=\psi.
\end{equation}
From Eqs. (\ref{E:sol1}) and (\ref{E:sol2}),
we then obtain $\nu'=2/r$ and thus
\begin{equation} \label{E:gtt}
   e^\nu  =C r^2,
\end{equation}
where $C$ is an integration constant.  Now from
Eq. (\ref{E:sol2}) we get
\[
   \xi^1 _{\phantom{1},1}=\frac{1}{2}
   (\psi'r+\psi).
\]
Substituting in Eq. (\ref{E:sol3}) and using
$\nu'=2/r$, simplification yields
\[
     \lambda'=-2\frac{\psi'}{\psi}.
\]
Finally, solving for $\lambda$, we have
\begin{equation}\label{E:grr}
   e^\lambda  = \left(\frac {B} {\psi}\right)^2,
\end{equation}
where $B$ is another integration constant.  When
substituting into Eqs.
(\ref{E:Einstein1})-(\ref{E:Einstein3}), it
becomes apparent that $B$ is merely a scale
factor, so that we may assume that $B=1$.  We
then get
\begin{equation}\label{E:lambda2}
   e^{-\lambda}=\psi^2
\end{equation}
and the Einstein field equations can be
rewritten as follows:
\begin{equation}\label{E:E1}
\frac{1}{r^2}\left(1 - \psi^2\right)
-\frac{(\psi^2)^{\prime}}{r}= 8\pi \rho,
\end{equation}
\begin{equation}\label{E:E2}
\frac{1}{r^2}\left(3\psi^2-1\right)= 8\pi p_r,
\end{equation}
and
\begin{equation}\label{E:E3}
\frac{\psi^2}{r^2}
+\frac{(\psi^2)^{\prime}}{r} =8\pi p_t.
\end{equation}
%END OF SECTION

\section{Wormhole structure}
Wormholes are handles or tunnels in spacetime
connecting widely separated regions of our
Universe or entirely different universes.  While
there were a number of forerunners, actual
physical structures suitable for interstellar
travel was first proposed by Morris and Thorne
\cite{MT88}.  Such wormholes can be described
by the static and spherically symmetric line
element
\begin{equation}\label{E:MT}
ds^{2}=-e^{2\Phi(r)}dt^{2}
  +\frac{dr^2}{1-b(r)/r}
+r^{2}(d\theta^{2}+\text{sin}^{2}\theta\,
  d\phi^{2}),
\end{equation}
using units in which $c=G=1$.  Here $\Phi =
\Phi(r)$ is called the \emph{redshift
function}, which must be everywhere finite
to avoid an event horizon.  The function
$b=b(r)$ is called the \emph{shape function}
since it determines the spatial shape of the
wormhole when viewed, for example, in an
embedding diagram \cite{MT88}.  The
spherical surface $r=r_0$ is the \emph
{throat} of the wormhole.  The shape
function must satisfy the following
conditions: $b(r_0)=r_0$, $b(r)<r$ for
$r>r_0$ and $b'(r_0)\le 1$, called the
\emph{flare-out condition}.  For a
Morris-Thorne wormhole, this condition
can only be satisfied by violating the
null energy condition, thereby becoming
the primary condition for the existence
of a traversable wormhole.

The discussion in Ref. \cite{MT88} was
based on the following strategy: specify
the geometric conditions required for a
traversable wormhole and then either
manufacture or search the Universe for
matter or fields that will produce the
corresponding energy-momentum tensor.
One of our goals in this paper is to
reverse this strategy by showing that
the conditions described are sufficient
for producing a complete solution, i.e.,
for obtaining both $\Phi=\Phi(r)$ and
$b=b(r)$, as well as the necessary
junction conditions.
%END OF SECTION

\section{Charged wormholes}\label{S:charged}

The motivation for a wormhole with a constant charge
$Q$, first proposed by Kim and Lee \cite{KL01}, was
 provided by the Reissner-Nordstr\"{o}m black hole
\begin{equation*}
  ds^2=-\left(1-\frac{2M}{r}+\frac{Q^2}{r^2}\right)dt^2
   +\left(1-\frac{2M}{r}+\frac{Q^2}{r^2}\right)^{-1}dr^2\\
   +r^2(d\theta^2+\text{sin}^2\theta\,d\phi^2),
\end{equation*}
suggesting that
\begin{equation}\label{E:lambda3}
   e^{\lambda(r)}=\left(1-\frac{b(r)}{r}
   +\frac{Q^2}{r^2}\right)^{-1}.
\end{equation}
Charged wormholes are also discussed in Refs.
\cite{pK11, pK16}.

Since we are assuming conformal symmetry, we
have $e^{\nu}=Cr^2$ from Eq. (\ref{E:gtt}).
Moreover, from Eq. (\ref{E:lambda2}),
\begin{equation}
   \psi^2=1-\frac{b(r)}{r}+\frac{Q^2}{r^2}.
\end{equation}
We also assume that $b=b(r)$ satisfies the
usual conditions for a shape function: letting
$r=r_1$ be the radius of the throat, we
require that $b(r_1)=r_1$ and $b(r_1)<1$,
while $b(r)<r$ for $r>r_1$.
%END OF SECTION

\section{Wormholes from a metric of embedding
class one}
Returning to Sec. \ref{S:embedding}, we know
from Eq. (\ref{E:lambda1}) that in view of
Eq. (\ref{E:lambda3})
\begin{equation}\label{E:lambda4}
   e^{\lambda(r)}=\frac{1}
   {1-\frac{b(r)}{r}+\frac{Q^2}{r^2}}=
   1+\frac{1}{4}Ke^{\nu(r)}[\nu'(r)]^2.
\end{equation}
Moreover, from Eq. (\ref{E:gtt}), $e^{\nu}=Cr^2$,
we have
\[
  1=\left(1-\frac{b(r)}{r}+\frac{Q^2}{r^2}
  \right)\left(1+\frac{1}{4}K(Cr^2)
     \frac{4}{r^2}\right)
\]
since $(\nu')^2=4/r^2$.  Hence
\[
   \frac{1}{1+KC}=1-\frac{b(r)}{r}
      +\frac{Q^2}{r^2}.
\]
(Observe that since $b(r_1)=r_1$, $Q^2$ cannot
be zero.)  Solving for $b(r)$, we get
\begin{equation}\label{E:shape1}
   b(r)=r\left(1+\frac{Q^2}{r^2}-
      \frac{1}{1+KC}\right).
\end{equation}
The condition $b(r_1)=r_1$ now leads to
$1+KC=r_1^2/Q^2$ and
\begin{equation}
   C=\frac{1}{K}\left(\frac{r_1^2}{Q^2}
   -1\right).
\end{equation}
The result is
\begin{equation}\label{E:shape2}
   b(r)=r\left(1+\frac{Q^2}{r^2}-
      \frac{Q^2}{r_1^2}\right).
\end{equation}
This result, in turn, leads to
\begin{equation}
   b'(r_1)=1-\frac{2Q^2}{r_1^2}<1,
\end{equation}
provided that $r_1^2>2Q^2$.

The flare-out condition is thereby satisfied,
but unlike a Morris-Thorne wormhole, satisfying
this condition does not automatically result
in a violation of the null energy condition
(NEC).  To see why, recall that the NEC
states that for the energy-momentum tensor
$T_{\alpha\beta}$,
$T_{\alpha\beta}k^{\alpha}k^{\beta}\ge 0$
for all null vectors $k^{\alpha}$.  Consider
the radial outgoing null vector $(1,1,0,0)$.
Then if $b=b(r)$ were the shape function of
a regular Morris-Thorne wormhole, Eq.
(\ref{E:MT}), we would have \cite{MT88}
\[
   8\pi\rho +8\pi p_r=\frac{b'(r)-b(r)/r}
   {r^2}+\frac{2\Phi'}{r}\left
      (1-\frac{b(r)}{r}\right).
\]
So at $r=r_1$, $8\pi\rho(r_1)+8\pi p_r(r_1)
<0$ since $b'(r_1)<1$.  The problem is that
our metric has been altered due to the
embedding, Eq. (\ref{E:lambda1}), i.e.,
$e^{\lambda(r)}=1+\frac{1}{4}Ke^{\nu(r)}
(\nu')^2$.  As a result, we are no longer
dealing with the same null vector: since
$b(r)$ has changed, so has $8\pi\rho(r_1)
+8\pi p_r(r_1)$.  On the other hand, we
also made use of Eq. (\ref{E:lambda3}),
$ e^{\lambda(r)}=(1-b(r)/r+Q^2/r^2)^{-1}$,
which gives us the \emph{effective shape
function} used in Ref. \cite{pK16}:
\begin{equation}
    b_{\text{eff}}(r)=b(r)-\frac{Q^2}{r}.
\end{equation}
This form suggests that Eq. (\ref{E:shape1})
be modified as follows:
\begin{equation}\label{E:shape3}
   b(r)=r\left(1+\frac{Q^2}{r^2}-
   \frac{1}{KC}\right)+\frac{Q^2}{r}.
\end{equation}
This modification needs to be justified by
showing that $b(r)$ in Eq. (\ref{E:shape3})
satisfies all the required conditions.  To
that end, let us denote the throat by
$r=r_0$, so that $b(r_0)=r_0$.  This
condition yields $1+KC=r_0^2/2Q^2$ and
\begin{equation}\label{E:constant}
   C=\frac{1}{K}\left(\frac{r_0^2}{2Q^2}
   -1\right).
\end{equation}
Substituting in Eq. (\ref{E:shape3}), we
get
\begin{equation}\label{E:shape4}
  b(r)=r\left(1+\frac{Q^2}{r^2}-
  \frac{2Q^2}{r_0^2}\right)+\frac{Q^2}{r}.
\end{equation}
Finally, the flare-out condition is also met:
\[
    b'(r_0)=1-\frac{4Q^2}{r_0^2}<1.
\]
(Since we want $b'(r_0)$ to be positive, we
also require that $r_0^2>4Q^2$.)

As noted earlier, we still need to check
the violation of the null energy condition:
From Eqs. (\ref{E:line1}), (\ref{E:lambda2}),
and (\ref{E:lambda3}),
\begin{equation*}
   e^{-\lambda(r)}=1-\frac{b(r)}{r}+
   \frac{Q^2}{r_0^2}=\psi^2(r).
\end{equation*}
Substituting $b(r)$, we get
\begin{equation}
  \psi^2(r)=1-\left(1+\frac{Q^2}{r^2}-
  \frac{2Q^2}{r_0^2}\right)-\frac{Q^2}{r^2}
  +\frac{Q^2}{r_0^2}=-\frac{2Q^2}{r^2}+
  \frac{3Q^2}{r_0^2}.
\end{equation}
Thus
\begin{equation}
  \psi^2(r_0)=\frac{Q^2}{r_0^2}.
\end{equation}
Also,
\begin{equation}
   (\psi^2(r))'=\frac{4Q^2}{r^3}.
\end{equation}
Returning now to the Einstein field
equations (\ref{E:E1}) and (\ref{E:E2}),
\begin{equation}
   \left. 8\pi (\rho+p_r)\right|_{r=r_0}
   =\left.\frac{1}{r^2}\left[2\psi^2(r)
   -\frac{(\psi^2(r))'}{r}
   \right]\right|_{r=r_0}
   =\frac{1}{r_0^2}\left(\frac{2Q^2}
   {r_0^2}\right)-\frac{4Q^2}{r_0^4}=
      -\frac{2Q^2}{r_0^4}<0.
\end{equation}
So the null energy condition is indeed
violated.

Returning now to Eq. (\ref{E:constant}), it is
clear that the free parameter $K$ can be used
to determine the constant $C$ and hence
$e^{\nu}$, which, in turn yields the redshift
function.

To complete the solution, we still need to
consider the following: we can see from Eq.
(\ref{E:gtt}), $e^{\nu}=Cr^2$, that our
wormhole spacetime is not asymptotically
flat.  So the wormhole material must be
cut off at some $r=a$ and joined to an
exterior Schwarzschild spacetime,
\begin{equation}
ds^{2}=-\left(1-\frac{2M}{r}\right)dt^{2}
+\frac{dr^2}{1-2M/r}
+r^{2}(d\theta^{2}+\text{sin}^{2}\theta\,
d\phi^{2}).
\end{equation}
Thus
\begin{equation}
   e^{\nu(a)}=Ca^2=1-\frac{2M}{a},
\end{equation}
where $M$ is the mass of the wormhole as
seen by a distant observer and $C$ is
obtained from Eq. (\ref{E:constant}).
It follows that the cut-off $r=a$ is
implicitly determined by the equation
\begin{equation}\label{E:junction}
   \frac{1}{K}\left(\frac{r_0^2}{2Q^2}
   -1\right)a^2=1-\frac{2M}{a}.
\end{equation}
So given $M$, $Q^2$, and $r_0$, the free
parameter $K$ determines the radius of
the junction surface.  Eq. (\ref
{E:junction}) will have a real solution
if $K$ is sufficiently large.
(Plausible values might be $10^9\,\
\text{m}^2$ and $10^{17}\,\text{m}^2$
arising in the discussion of compact
stellar objects in Ref. \cite{MDRK}.)
%END OF SECTION

\section{Flat galactic rotation curves}

One goal in many modified gravitational
theories is to explain the peculiar
behavior of galactic rotation curves
without postulating the existence of
dark matter, whether this be
noncommutative geometry \cite{fR12,
KG14} or $f(R)$ modified gravity
\cite{BHL08}.  The basic problem is
that test particles move with constant
tangential velocity $v^{\phi}$ in a
circular path sufficiently far from
the galactic core.  Taking the
observed rotation curves as input,
it is well known that
\begin{equation}\label{E:flat}
   e^{\nu}=B_0r^l,
\end{equation}
where $l=2v^{2\phi}$ and $B_0$ ia
an integration constant \cite{NVM09}.
In addition, it is shown in Ref.
\cite{MGN00} that in the presumed
dark-matter dominated region,
$v^{\phi}\approx 300\,\,
\text{km}/\text{s}=10^{-3}$ for a
typical galaxy.  So $l=0.000001$
\cite{kK09}.

The existence of flat rotation curves
indicates that the matter in the
galaxy increases linearly in the
outward radial direction.  To recall
the reason for this, suppose $m_1$
is the mass of a star, $v^{\phi}$ the
constant tangential velocity, and
$m_2$ the mass of everything else.
Multiplying $m_1$ by the centripetal
acceleration yields
\begin{equation}
   m_1\frac{v^{2\phi}}{r}=m_1m_2
   \frac{G}{r^2},
\end{equation}
where $G$ is Newton's gravitational
constant.  The result is (since $G=1$)
\begin{equation}\label{E:mass}
   m_2=rv^{2\phi}.
\end{equation}

Eq. (\ref{E:mass}) essentially
characterizes the dark-matter
hypothesis, but, as noted above,
other explanations are possible.
As a starting point, suppose we
consider \cite{MTW}
\[
   e^{\lambda(r)}=
   \frac{1}{1-2m(r)/r}.
\]
Then $m(r)=\frac{1}{2}r(1-
e^{-\lambda})$.  Since $e^{-\lambda}
\rightarrow 1$ as $r\rightarrow
\infty$, we can assume that $m(r)$
ia approximately constant over a
large range of $r$.  In other words,
$m(r)=Cr$ for some constant $C$.
Unfortunately, by Eq. (\ref{E:mass}),
$C$ has to be approximately equal to
$v^{2\phi}$.  This obstacle can be
overcome by the embedding theory in
Sec. \ref{S:embedding}.  Using Eq.
(\ref{E:lambda1}),
\begin{equation}\label{E:mass2}
   m(r)=\frac{1}{2}r\left(1-
   \frac{1}{1+\frac{1}{4}Ke^{\nu}
   (\nu')^2}\right),
\end{equation}
the free parameter $K$ gives us the
extra degree of freedom to produce the
correct values, provided, of course,
that $e^{\nu}$ and $(\nu')^2$ are
indeed approximately constant.  To
that end,  let us return to Refs.
\cite{NVM09}-\cite{kK09}, which deal
with typical galaxies, including our
own.  Suppose we assume for now that
$B_0=1$ in Eq. (\ref{E:flat}).  So
with our own galaxy in mind, let us
consider the range from 8 kps to
50 kps, associated with flat galactic
rotation curves.  Then $r^l$ ranges from
\[
   (26\,000\times 9.46\times 10^{15})
   ^{0.000001}\approx 1.000047
\]
to
\[
    (6.25\times 26\,000\times 9.46
    \times 10^{15})^{0.000001}\approx
    1.000049.
\]
These calculations show that the value
of $B_0$ has little effect, so that
$e^{\nu}$ does remain approximately
constant.

The values for $(\nu')^2$ are much
less robust.  However, of great help in
this situation is that $B_0$ drops out
entirely: from $e^{\nu}=B_0r^l$,
we have
\[
    (\nu')^2=\frac{l^2}{r^2}.
\]
So we can obtain an adequate approximation
for the above range, while in the range
16 kps to 30 kps, the resulting values
for $(\nu')^2$ are essentially fixed:
$4.1\times 10^{-54}\,\text{m}^{-2}$
and $1.2\times 10^{-54}\,\text{m}^{-2}$,
respectively.  For these values, $K
\approx 10^{48}\,\text{m}^2$.

With this choice of $K$, Eq.
(\ref{E:mass2}) reduces to $m(r)=
v^{2\phi}r$, thereby producing another
alternative to the dark-matter
hypothesis.  This outcome may be
viewed as the analogue of the
induced-matter theory in Ref.
\cite{WP92}, i.e., one could
maintain that the five-dimensional
flat spacetime impinges on our
Universe to produce the effect
that we normally interpret as
dark matter.

\emph{Remark:} The existence of two models
from the same embedding theory invites the
following speculation: according to
Brownstein and Moffet \cite{BM07}, a
significant amount of dark matter is
missing in the Bullet Cluster
1E0657-558.  This is also the cluster
that has supposedly shown that dark
matter actually exists.  The main argument
in Ref. \cite{BM07} is that this phenomenon
can be explained by means of a modified
gravitational theory, to which we could
add the present embedding theory.  On the
other hand, if the dark-matter hypothesis
is to be retained and if some of the dark
matter is indeed missing, then the
existence of a conformally symmetric
charged wormhole may be the preferred
explanation: the Bullet Cluster consists
of two colliding galaxies moving at very
high velocities, so that the dark matter
could be literally driven into the
wormhole.
%END OF SECTION

\section{Conclusion}
It is well known that a curved spacetime
can be embedded in a higher-dimensional
flat spacetime.  A spacetime is said to
be of class $m$ if $n+m$ is the lowest
dimension of the flat space in which the
given space can be embedded.  Following
Ref. \cite{MG17}, we assume a spherically
symmetric metric of class two that can
be reduced to class one by a suitable
transformation.

These ideas have been applied to two
completely different models, a new solution
for a charged wormhole admitting a
one-parameter group of conformal motions
and a new model to explain the flat
rotation curves in spiral galaxies without
the need for dark matter.  The existence
of the latter model can be attributed to
the free parameter $K$ in the embedding
theory.  In the former case, the free
parameter $K$ plays an equally critical
role in obtaining a complete wormhole
solution: $K$ helps determine the
redshift and shape functions, as well
as the radius of the junction interface
that joins the interior solution to an
exterior Schwarzschild spacetime.


\begin{thebibliography}{30}
\bibitem{WP92}P.S. Wesson, J. Ponce de Le\'{o}n, J. Math. Phys.
   33 (1992) 3883.
\bibitem{MG17}S.K. Maurya, M. Govender, Eur. Phys. J. C
   77 (2017) 347.
\bibitem{G1}S.K. Maurya, Y.K. Gupta, S. Ray, D. Deb, Eur.
   Phys. J. C 77 (2017) 45.
\bibitem{G2}S.K. Maurya, Y.K. Gupta, S. Ray, D. Deb, Eur.
   Phys. J. C 76 (2016) 693.
\bibitem{MDRK}S.K. Maurya, D. Deb, S. Ray, P.K.F. Kuhfittig,
    arXiv: 1703.08436.
 \bibitem{kK48}K.R. Karmarkar, Proc. Ind. Acad. Sci. 27
    (1948) 56.
 \bibitem{pB16}P. Bhar, S.K. Maurya, Y.K. Gupta, T. Manna,
    Eur. Phys. J. A 52 (2016) 312.
\bibitem{MM96}R. Maartens, C.M. Mellin, Class. Quantum Grav.
    13 (1996) 1571.
\bibitem{BHL07}C.G. B\"{o}hmer, T. Harko, F.S.N. Lobo,
    Phys. Rev. D 76 (2007) 084014.
\bibitem{HPa}L. Herrera, J. Ponce de Le\'{o}n, J. Math. Phys.
   26 (1985) 778.
\bibitem{HPb}L. Herrera, J. Ponce de Le\'{o}n, J. Math. Phys.
   26 (1985) 2018.
\bibitem{MS93}M. Mars, J.M.M. Senovilla, Class. Quantum Grav.
   10 (1993) 1633.
\bibitem{Ray08}S. Ray, A.A. Usmani, M. Kalam, K. Chakraborty,
   Indian J. Phys. 82 (2008) 1191.
\bibitem{fR10}F. Rahaman, M. Jamil, M. Kalam, K. Chakraborty,
   A. Ghosh, Astrophys. Space Sci. 325 (2010) 137.
\bibitem{fR12b}F. Rahaman, S. Ray, I. Karar, H.I. Fatima,
   S. Bhowmick, G.K. Ghosh, arxiv: 1211.1228 [gr-qc].
\bibitem{BHL08a}C.G. B\"{o}hmer, T. Harko, F.S.N. Lobo,
   Class. Quantum Grav. 25 (2008) 075016.
\bibitem{CMS10}A. Castro, A. Maloney, A. Strominger,
   Phys. Rev. D 82 (2010) 024008.
\bibitem{RRKKI}F. Rahaman, S. Ray, G.S. Khadekar, P.K.F.
   Kuhfittig, I. Karar, Int. J. Theor. Phys. 54 (2015)
   699.
\bibitem{pK10}P.K.F. Kuhfittig, Indian J. Phys. 90
   (2010) 877.
\bibitem{MT88}M.S. Morris, K.S. Thorne, Amer. J. Phys.
   56 (1988) 395.
\bibitem{KL01}S.-W. Kim, H. Lee, Phys. Rev. D 63 (2001)
   064014.
\bibitem{pK11}P.K.F. Kuhfittig, Central Eur. J. Phys.
   9 (2011) 144.
\bibitem{pK16}P.K.F. Kuhfittig, J. Appl. Math. Phys.
   (JAMP) 4 (2016) 2117.
\bibitem{fR12}R. Rahaman, P.K.F. Kuhfittig, K.
   Chakraborty, A.A. Uamani, S. Ray, Gen. Rel. Gravit.
   44 (2012) 905.
\bibitem{KG14}P.K.F. Kuhfittig, V.D. Gladney, J. Mod.
   Phys. 5 (2014) 1931.
\bibitem{BHL08}C.G.  B\"{o}hmer, T. Harko, F.S.N. Lobo,
   Astropart. Phys. 29 (2008) 386.
\bibitem{NVM09}K.K. Nandi, I. Valitov, N.G. Migranov,
   Phys. Rev. D 62 (2009) 047301.
\bibitem{MGN00}T. Matos, F.S. Guzman, D. Nunez, Phys.
   Rev. D 62 (2008) 061301R.
\bibitem{kK09}K.K. Nandi, A.I. Filippov, F. Rahaman,
   S. Ray, A.A. Usmani, M. Kalam, A. DeBenedictis,
   Mon. Not. R. Astron. Soc. 399 (2009) 2079.
\bibitem{MTW}C.W. Misner, K.S. Thorne, J.A. Wheeler,
   Gravitation, W. Freeman and Company, New York,
   1973, page 608.
\bibitem{BM07}J.R. Brownstein, J.W. Moffet, Mon.
   Not. Roy. Astron. Soc. 382 (2007) 29.
\end{thebibliography}
\end{document}